\documentclass[onecolumn,showpacs,preprintnumbers,amsmath,amssymb,showkeys,11pt]{revtex4}

\DeclareMathOperator{\Spur}{Tr}
\newcommand{\ens}[0]{\ensuremath} %Kurzschreibweise für mathematische Definitionen
\newcommand{\anfEngl}[1]{``#1''} %englische Anführungszeichen
\newcommand{\SkM}[2]{\ens{\left\langle#1,#2\right\rangle}} %Skalarprodukt in mathematischer Notation
\newcommand{\x}[0]{\ens{\otimes}} %Tensorprodukt
\newcommand{\isom}[0]{\ens{\cong}} %isomorph
\newcommand{\nach}[0]{\ens{\rightarrow}} %Mengenabbildung
\newcommand{\Mg}[1]{\ens{\left\lbrace #1 \right\rbrace}} %Def. nur wg. der Klammern
\newcommand{\MgE}[1]{\ens{\Mg{1,\dots,#1}}} % Mengen der Form {1,...,#1}
\newcommand{\Folge}[3]{\ens{(#1_#2)_{#2 \in #3}}}
\newcommand{\Fkt}[3]{\ens{#1 : #2 \nach #3}} %Funktion : z. B. "f : X -> Y"
\def\idty{{\leavevmode{\rm 1\ifmmode\mkern -4.4mu\else\kern -.3em\fi I}}}
\newcommand{\Eins}[0]{\ens{\idty}} %Einheitsoperator
\newcommand{\Proof}[0]{\emph{Proof: }} %Beginn eines Beweises
\newcommand{\BE}[0]{\hfill $\Box$} %Beweisende
\newcommand{\cH}[0]{\ens{\mathcal{H}}}
\newcommand{\cJ}[0]{\ens{\mathcal{J}}}
\newcommand{\fA}[0]{\ens{\mathfrak{A}}}
\newcommand{\fB}[0]{\ens{\mathfrak{B}}}

\newcommand{\N}[0]{\ens{\mathbb{N}}} %Definitionen der Mengen
\newcommand{\C}[0]{\ens{\mathbb{C}}}
\newcommand{\iE}[0]{\ens{\mathrm{i}}} %imaginäre Einheit
\newtheorem{Theorem}{Theorem}
\newtheorem{Lemma}{Lemma}
\newtheorem{Corollary}{Corollary}

\begin{document}
\title{The Jamio{\l}kowski isomorphism and a conceptionally simple proof for the correspondence between vectors
  having Schmidt number $k$ and $k$-positive maps}
\author{Kedar S. Ranade}\author{Mazhar Ali}
\affiliation{Institut f\"ur Angewandte Physik, Technische Universit\"at Darmstadt, Hochschulstra{\ss}e 4a,
  64289 Darmstadt, Deutschland/Germany}
\date{February $27^{\mathrm{th}}$, 2007}


\begin{abstract}
  \noindent Positive maps which are not completely positive are used in quantum information theory
  as witnesses for convex sets of states, in particular as entanglement witnesses and more generally
  as witnesses for states having Schmidt number not greater than $k$. It is known that such witnesses
  are related to $k$-positive maps. In this article we propose a new proof for the correspondence between
  vectors having Schmidt number $k$ and $k$-positive maps using Jamio{\l}kowski's criterion for positivity
  of linear maps; to this aim, we also investigate the precise notion of the term \anfEngl{Jamio{\l}kowski isomorphism}.
  As consequences of our proof we get the Jamio{\l}kowski criterion for complete positivity,
  and we find a special case of a result by Choi, namely that $k$-positivity implies complete positivity, if $k$
  is the dimension of the smaller one of the Hilbert spaces on which the operators act.
\end{abstract}

\maketitle

\section{Introduction}\thispagestyle{plain}
\noindent In quantum mechanics, the state of a physical system is described by a density operator $\rho$, which is a
positive semi-definite operator (i.\,e. all eigenvalues of $\rho$ are non-negative: $\rho \geq 0$) acting on some
Hilbert space associated with that physical system. When a physical system undergoes a dynamical process, then at
the end of that process the new state of the same physical system should also be represented by a valid density operator
$\rho^\prime$; this means that the dynamical process is represented by a positive map $T$, such that
$\rho^\prime = T(\rho)$. It was observed that this condition only cannot sufficiently describe physical
processes \cite{GL75}: For any state $\rho_{AB}$ of a joint physical system~$AB$, where the subsystems $A$ and~$B$
are spatially separated and each subsystem is transformed according to positive maps $T_A$ and~$T_B$, respectively,
the global state must still be represented by a valid density operator. This leads to the condition that the tensor
product of these maps, $T = T_A \x T_B$, must be a positive map. Thus, the maps $T_A$ and $T_B$ have
to be completely positive in the following sense: \emph{A linear map $T$ is completely positive, if and only if
$\Eins_k \x T$ is positive for all $k \in \N$.} Here, $\Eins_k$ denotes the identity map acting on the set of
$k \times k$-matrices.
\par The structure of completely positive maps and their applications in quantum theory were \mbox{extensively} studied
in the 1960s and 70s. However, positive maps which are not completely positive are also known to have applications
in quantum information theory. In particular, they can be used as so-called \anfEngl{entanglement witnesses}.
The simplest example of such a map is the transposition $T_\mathrm{tr}$: if $\Eins_A \x T_{\mathrm{tr},B}(\rho_{AB})$
turns out to be non-positive, then $\rho_{AB}$ is entangled \cite{P96}.
This work has started considerable efforts to understand the positive maps which are not completely
positive~\cite{HTZ05}. It was also observed that there is a direct relation between positive maps and entanglement witnesses through the so-called Jamio{\l}kowski isomorphism; the Peres criterion \cite{P96} and the reduction criterion
for separability \cite{H99,C99} are manifestations of this relation between positive maps and entanglement
witnesses \cite{C06}.
\par Terhal and Horodecki \cite{Ter} extended the notion of the Schmidt number for pure states to mixed states, where
it turned out to be a legitimate measure of entanglement. If we denote by $S_k$ the convex hull of all pure
states having Schmidt number not greater than $k$, then the set of separable states is $S_1$, whereas all other
states are entangled. The states in these sets can be characterised with the aid of
\emph{Schmidt witnesses} \cite{SBL01}: A hermitian operator $W$ is called Schmidt witness of class~$k$, if and
only if $\Spur(W \sigma)\geq 0$ for all $\sigma \in S_{k-1}$ and there exists at least one $\rho \in S_k$ such that
$\Spur(W \rho) < 0$.
\par Terhal and Horodecki \cite{Ter} also presented a result relating $k$-positivity to Schmidt witnesses of class $k$.
As Clarisse \cite{Cla} noted, their result can be stated as follows: \textit{A map is $k$-positive, if and only if the
corresponding operator [under the so-called Jamio{\l}kowski isomorphism] is positive on states with Schmidt number $k$
or less.} We will present a conceptually simple, direct and explicit proof of that fact; although the basic idea of
the proof is quite simple, it does not seem to have been published in the literature. To this aim, we discuss the
fact that there are actually two \anfEngl{Jamio{\l}kowski isomorphisms}, which are related, but nevertheless
have different properties. We show, that in the part of the work by de Pillis \cite{Pil} and Jamio{\l}kowski \cite{Jam}
which is relevant for us, either isomorphism may have been used, whereas for the above statement only one of these
isomorphisms is appropriate.
\par This work is organized as follows: In section \ref{Prelim} we introduce the necessary notation and shortly discuss
the notion of $k$-positivity and the Schmidt decomposition. In sections \ref{JamIso1} and \ref{JamIso2} we discuss
the main properties of the two Jamio{\l}kowski isomorphisms. In section \ref{MainSec} we state and prove the relation
between $k$-positive maps and vectors having Schmidt number $k$; we conclude the section with a further discussion of
the two Jamio{\l}kowski isomorphisms and note some consequences of the main theorem. Finally we summarize our
results in section \ref{Summary}.

\section{Preliminaries}\label{Prelim}
\noindent Following de Pillis \cite{Pil} and Jamio{\l}kowski \cite{Jam} we denote by $\C$ the field of complex
numbers, the complex conjugate of $z \in \C$ by $\overline{z}$ and the transpose and adjoint of an operator $A$
by $A^t$ and $A^*$, respectively. Scalar products are denoted by $\SkM{\,\cdot\,}{\,\cdot\,}$ and are taken to be
linear in the left and antilinear in the right argument; occasionally we indicate the space on which they are taken
by a subscript.
\par We consider two finite-dimensional Hilbert spaces $\cH_A := \C^n$ and $\cH_B := \C^m$,
to which there are assigned their respective algebras $\fA := \C^{n \times n}$ and $\fB := \C^{m \times m}$ of
matrices acting on $\cH_A$ and $\cH_B$, which themselves form Hilbert spaces using the Hilbert-Schmidt scalar
product, i.\,e. $\SkM{A_1}{A_2} = \Spur A_2^*A_1$ for $A_1,\,A_2 \in \fA$ and similarly for $\fB$.
\par The space of linear maps from $\fA$ to $\fB$ is denoted by $L(\fA,\fB)$. A map $T \in L(\fA,\fB)$ is called
\emph{hermiticity-preserving}, if it maps hermitian $A \in \fA$ to hermitian $T(A) \in \fB$; it is called \emph{positive},
if it maps positive $A$ to positive $T(A)$. For any $k \in \N$ a map $T \in L(\fA,\fB)$ gives rise to a map
$T_k := \Eins_k \x T \in L(M_k(\fA),M_k(\fB))$, where $M_k(\fA)$ and $M_k(\fB)$ are the sets of matrices having
elements of $\fA$ and $\fB$ as their entries; for a matrix $A = (a_{ij})_{i,j = 1}^{k}$, where $a_{ij} \in \fA$,
it is defined by
\begin{equation}
  T_k\left[\begin{pmatrix}
    a_{11} & \hdots & a_{1k}\\ \vdots & \ddots & \vdots\\ a_{k1} & \hdots & a_{kk}
 \end{pmatrix} \right] := \begin{pmatrix}
    T(a_{11}) & \hdots & T(a_{1k})\\ \vdots & \ddots & \vdots\\ T(a_{k1}) & \hdots & T(a_{kk})
 \end{pmatrix}.
\end{equation}
The map $T$ is called \emph{$k$-positive}, if $T_k$ is positive; it is called \emph{completely positive}, if
it is $k$-positive for all $k \in \N$. A detailed mathematical discussion of these properties may be found
in a paper by Choi \cite{Ch}.
\par Physically, $T_k$ may be understood as coupling an auxiliary system (an \anfEngl{ancilla}) of dimension~$k$
to a quantum system without performing any action on that ancilla. Thus, if the use of such ancilla is allowed
and if arbitrary quantum states of the principal system can be prepared, an operation consistent with quantum
mechanics must be $k$-positive, and in the general case of arbitrary dimension of the ancilla, it must be completely
positive.
\par We will prove a theorem relating the $k$-positivity of a map $T \in L(\fA,\fB)$ to vectors on
$\cH_A \x \cH_B$ having Schmidt number $k$ or less. For this purpose, we will now state a well-known result known as
the \emph{Schmidt decomposition} of a vector in $\cH_A \x \cH_B$. A proof of this result may be found e.\,g.
in the book by Nielsen and Chuang \cite{NC}.\pagebreak
\begin{Theorem}[Schmidt decomposition]\label{Schmidt}\hfill\\
  For each non-zero vector $v \in \cH_A \x \cH_B$ there exists a number $s \in \MgE{\min\Mg{n,m}}$,
  positive numbers $\lambda_1, \dots, \lambda_s$ and orthonormal systems $(e_i^A)_{i = 1}^{s}$ and
  $(e_i^B)_{i = 1}^{s}$ in $\cH_A$ and $\cH_B$, such that
  \begin{equation*}
    v = \sum\nolimits_{i = 1}^{s} \lambda_i e_i^A \x e_i^B
  \end{equation*}
  holds.
\end{Theorem}
The number $s$ is the rank of either reduced density matrix, thus well-defined; it is called the
\emph{Schmidt number} of the vector $v$. For later use in our proof, we note the following:
\begin{Lemma}[Schmidt number of certain vectors]\label{JamSchmidt}\hfill\\
  Let $(v_i^A)_{i = 1}^s$ and $(v_i^B)_{i = 1}^s$ be systems of not necessarily orthogonal vectors
  in Hilbert spaces $\cH_A$ and~$\cH_B$. Then, the Schmidt number of the vector
  $v := \sum_{i = 1}^{s} v_i^A \x v_i^B$ is not greater than $s$.
\end{Lemma}
\Proof The Hilbert spaces spanned by the vector systems $(v_i^A)_{i = 1}^s$ and $(v_i^B)_{i = 1}^s$ have
dimension not greater than $s$. Thus, by Theorem \ref{Schmidt} the Schmidt number of $v$ cannot be greater
than $s$. \BE

\section{The Jamio{\l}kowski isomorphism and its properties}\label{JamIso1}
\noindent De Pillis \cite{Pil} considered a mapping $\Fkt{\cJ_1}{L(\fA,\fB)}{\fA \x \fB}$ which has the defining
property that $\SkM{\cJ_1(T)}{A^* \x B}_{\fA \x \fB} = \SkM{T(A)}{B}_\fB$ should hold for all $T \in L(\fA,\fB)$,
$A \in \fA$ and $B \in \fB$. He proved the following properties of such a map:
\begin{Lemma}[Basic properties of $\cJ_1$]\label{Jam}\hfill\\
  The mapping $\cJ_1$ is uniquely defined, and for any orthonormal basis $\Folge{E}{i}{I}$ of $\fA$
  and every operator $T \in \fA$ the equation $\cJ_1(T) = \sum_{i \in I} E_i^* \x T(E_i)$ holds.
  Furthermore, $\cJ_1$ is an isometric isomorphism of the Hilbert spaces $L(\fA,\fB)$ and $\fA \x \fB$.
\end{Lemma}
The following Lemma characterises hermiticity-preserving maps; it was used by de Pillis \cite{Pil}.
\begin{Lemma}[Hermiticity-preserving maps]\label{Pillis-hilf}\hfill\\
  A linear map $T \in L(\fA,\fB)$ preserves hermiticity, if and only if $T(A^*) = T(A)^*$ holds for all
  $A \in \fA$. If $\Folge{E}{i}{I}$ is a basis of $\fA$, this is the case, if and only if
  $T(E_i^*) = T(E_i)^*$ for all $i \in I$.
\end{Lemma}
\Proof If $T$ preserves hermiticity, we have $T(A)^* = T(A) = T(A^*)$ for hermitian $A$; in general,
an arbitrary $A \in \fA$ may be decomposed into $A = A_1 + \iE A_2$, where $A_1,\,A_2 \in \fA$ are hermitian,
so that
\begin{equation}
  T(A^*) = T(A_1^* - \iE A_2^*)    = T(A_1^*) - \iE T(A_2^*)
         = T(A_1)^* - \iE T(A_2)^* = [T(A_1) + \iE T(A_2)]^* = T(A)^*
\end{equation}
holds. On the converse, we calculate $T(A) = T(A^*) = T(A)^*$ for hermitian $A \in \fA$, which shows the
first statement. For the second statement, we use the decomposition $A = \sum_i a_i E_i$ and
calculate
\begin{equation}
  T(A^*) = \sum\nolimits_{i \in I} \overline{a_i}\,T(E_i^*)
    \stackrel{\text{by assumption}}{=} \sum\nolimits_{i \in I} \overline{a_i}\,T(E_i)^* = T(A)^*.
\end{equation}
The inverse statement is obvious, and this concludes the proof. \BE
\begin{Theorem}[Maps which preserve hermiticity and/or positivity I]\label{J1Krit}\hfill\\
  A linear map $T \in L(\fA,\fB)$ is
  \begin{itemize}
    \item hermiticity-preserving, if and only if $\cJ_1(T)$ is hermitian.
    \item positive, if and only if $\SkM{\cJ_1(T)x \x y}{x \x y} \geq 0$ holds for all $x \in \cH_A$
      and all $y \in \cH_B$.
  \end{itemize}
\end{Theorem}
The first part was proved by de Pillis \cite{Pil}, the second by Jamio{\l}kowski \cite{Jam}; the latter
criterion can be interpreted as follows: \textit{A map $T$ is positive, if and only if $\cJ_1(T)$ is positive
on separable vectors}. We will use a modified version of Jamio{\l}kowski's criterion, which is appropriate
for our proof.

\section{The modified Jamio{\l}kowski isomorphism}\label{JamIso2}
\noindent In this section we focus on a particular basis of $\fA$, namely the basis $(E_{ij})_{i,j = 1}^{n}$
which consists of matrices $E_{ij}$ which have entry one in the $j$-th column of the $i$-th row, whereas all
other entries are zero; this basis is sometimes called \emph{Weyl basis}. It has the property
that $E_{ij} = \overline{E_{ij}} = E_{ji}^* = E_{ji}^t$ holds for all $i,\,j \in \MgE{n}$.
\par We now consider a variant of the Jamio{\l}kowski isomorphism, which we will call $\cJ_2$ and which
is defined by (cf. e.\,g. \cite{Sal,KW})
\begin{equation}
  \cJ_2(T) := \sum\nolimits_{i,j = 1}^{n} E_{ij} \x T(E_{ij}).
\end{equation}
The difference between $\cJ_1$ and $\cJ_2$ is that in the first tensor factor there is no adjoint. Note that
for $T_1,T_2 \in L(\fA,\fB)$, due to the fact that
$\SkM{E_{ij}^*}{E_{kl}^*} = \delta_{ik} \delta_{jl} = \SkM{E_{ij}}{E_{kl}}$, we have
\begin{align}
  \SkM{\cJ_1(T_1)}{\cJ_1(T_2)}
    &= \SkM{\sum\nolimits_{ij} E_{ij}^* \x T_1(E_{ij})}{\sum\nolimits_{kl} E_{kl}^* \x T_2(E_{kl})}\\
    &= \sum\nolimits_{ijkl} \SkM{E_{ij}^*}{E_{kl}^*} \SkM{T_1(E_{ij})}{T_2(E_{kl})}
    = \sum\nolimits_{ijkl} \SkM{E_{ij}}{E_{kl}} \SkM{T_1(E_{ij})}{T_2(E_{kl})}\\
    &= \SkM{\sum\nolimits_{ij} E_{ij} \x T_1(E_{ij})}{\sum\nolimits_{kl} E_{kl} \x T_2(E_{kl})}
    = \SkM{\cJ_2(T_1)}{\cJ_2(T_2)}.
\end{align}
This shows that $\cJ_2$ is indeed an isomorphism, and we can adapt Theorem \ref{J1Krit} to this modified isomorphism.
\begin{Lemma}[Maps which preserve hermiticity and/or positivity II]\label{JamVar}\hfill\\
  A linear map $T \in L(\fA,\fB)$ is
  \begin{itemize}
    \item hermiticity-preserving, if and only if $\cJ_2(T)$ is hermitian.
    \item positive, if and only if $\SkM{\cJ_2(T)x \x y}{x \x y} \geq 0$ holds for all $x \in \cH_A$
      and all $y \in \cH_B$.
  \end{itemize}
\end{Lemma}
\Proof For showing the first statement, one can calculate
\begin{align}
  \cJ_2(T)^* &= \left(\sum\nolimits_{ij} E_{ij} \x T(E_{ij})\right)^* = \sum\nolimits_{ij} E_{ij}^* \x T(E_{ij})^*\\
  &\stackrel{?}{=} \sum\nolimits_{ij} E_{ji} \x T(E_{ij}^*) = \sum\nolimits_{ij} E_{ji} \x T(E_{ji}) = \cJ_2(T),
\end{align}
where the equality in question holds, if and only if $T$ preserves hermiticity (according to Lemma~\ref{Pillis-hilf}).
The proof of the second statement is nearly the same as Jamio{\l}kowski's original proof \cite{Jam}: As any positive
operator $T$ may be decomposed into a real linear combination of projection operators, we have to show that
$T(P_x)$ is positive for any one-dimensional projection $P_x$ projecting on the vector space spanned by some
unit vector $x \in \cH_A$. Using an orthonormal basis $(f_p)_{p = 1}^{n}$ of $\cH_A$, we therefore calculate
\begin{align}
  T(P_x) &= \sum\nolimits_{ij} \SkM{P_x}{E_{ij}}_\fA T(E_{ij}) = \sum\nolimits_{ij} \Spur(E_{ij}^* P_x) T(E_{ij})\\
         &= \sum\nolimits_{ijp} \SkM{E_{ji}P_x f_p}{f_p}_{\cH_A} T(E_{ij})
          = \sum\nolimits_{ijp} \SkM{E_{ji}x}{f_p}_{\cH_A} \SkM{f_p}{x}_{\cH_A} T(E_{ij})\\
         &= \sum\nolimits_{ij} \SkM{E_{ji}x}{x}_{\cH_A} T(E_{ij}).
\end{align}
Thus, $T(P_x)$ is positive, if and only if $\sum_{ij} \SkM{E_{ji}x}{x}_{\cH_A} \SkM{T(E_{ij})y}{y}_{\cH_B} \geq 0$
for all $x \in \cH_A$ and all $y \in \cH_B$. If we use $x = (x_1, \dots, x_n) \in \cH_A$ and its complex conjugate
$\overline{x} = (\overline{x_1}, \dots, \overline{x_n}) \in \cH_A$, we get
$\SkM{E_{ji}x}{x} = x_i \cdot \overline{x_j} = \SkM{E_{ij}\overline{x}}{\overline{x}}$, that is
\begin{equation}
  \sum\nolimits_{ij} \SkM{E_{ji}x}{x}_{\cH_A} \SkM{T(E_{ij})y}{y}_{\cH_B}
  = \sum\nolimits_{ij} \SkM{E_{ij}\overline{x}}{\overline{x}}_{\cH_A} \SkM{T(E_{ij})y}{y}_{\cH_B}
  = \SkM{\cJ_2(T) \overline{x} \x y}{\overline{x} \x y},
\end{equation}
which has to hold for all $x \in \cH_A$ and all $y \in \cH_B$. Since $x \in \cH_A$ implies $\overline{x} \in \cH_A$
(for $\cH_A = \C^n$) and vice versa, the Lemma is proved. \BE\\
A notable difference between the isomorphisms $\cJ_1$ and $\cJ_2$ is that $\cJ_1$ does not depend upon a particular
basis of $\fA$, whereas $\cJ_2$ does: consider another basis $(F_{ij})_{i,j = 1}^{n}$ of $\fA$,
which may be expressed as $F_{ij} = \sum_{kl} \SkM{F_{ij}}{E_{kl}}_\fA E_{kl}$ for $i,\,j \in \MgE{n}$. Using this basis
to define $\cJ_2^\prime$ we find $\cJ_2^\prime(T) = \sum_{klpq} \left[\sum_{ij} \SkM{F_{ij}}{E_{kl}}
\SkM{F_{ij}}{E_{pq}}\right] E_{kl} \x T(E_{pq})$, and this is equal to $\cJ_2(T)$ for all $T$, if and only if the
inner bracket is $\delta_{kp} \delta_{lq}$. As an example, consider the canonical basis $(e_i)_{i = 1}^{n}$ of $\cH_A$
and a unitary operator $U$ on $\cH_A$; thus, the vectors $f_i := Ue_i$ also form an orthonormal basis of $\cH_A$.
Defining $F_{ij} := \SkM{\,\cdot\,}{f_j} f_i$, it may be shown that $\cJ_2 = \cJ_2^\prime$, if and only if $U$ is
orthogonal in the sense, that $U^tU = \Eins$. Since there are unitary, but non-orthogonal transformations, $\cJ_2$
is basis-dependent.

\section{The main theorem and its consequences}\label{MainSec}
\noindent We will now state the main theorem and prove it using the material from the previous section.
\begin{Theorem}[Main theorem]\label{Main}\hfill\\
  A linear mapping $T \in L(\fA,\fB)$ is $k$-positive, if and only if the inequality $\SkM{\cJ_2(T)v}{v} \geq 0$
  holds for all vectors $v \in \cH_A \x \cH_B$ having Schmidt number less or equal than $k$.
\end{Theorem}
The basic idea of the proof is quite simple: by definition, $T$ is $k$-positive, if and only if the operator
$T_k := \Eins_k \x T \in L(M_k(\fA),M_k(\fB))$ is positive. The latter property can be checked by using the modified 
Jamio{\l}kowski criterion of Lemma \ref{JamVar}. This will be formalized in the following proof.\\
\Proof Let $\cH_{A;k} := \C^k \x \cH_A \isom \oplus_{\alpha = 1}^{k} \cH_A$ and $\cH_{B;k} := \C^k \x \cH_B
\isom \oplus_{\beta = 1}^{k} \cH_B$ be two Hilbert spaces with associated matrix algebras $\fA_k := M_k(\fA)$ and
$\fB_k := M_k(\fB)$. The operator $T_k$ is positive, if for all $x \in \cH_{A;k}$ and $y \in \cH_{B;k}$ the inequality
$\SkM{\cJ_{2;k}(T_k)x \x y}{x \x y} \geq 0$  holds, where $\Fkt{\cJ_{2;k}}{L(\fA_k,\fB_k)}{\fA_k \x \fB_k}$
denotes the modified Jamio{\l}kowski isomorphisms on the respective spaces.
\par Considering the orthonormal basis $(E_{ij})_{i,j = 1}^{n}$ of $\fA$ and $(e_{\alpha\beta})_{\alpha,\beta=1}^{k}$
of $M_k(\C)$, we have the ortho\-normal basis $(e_{\alpha\beta} \x E_{ij})_{i,j = 1}^{n}{}_{\alpha,\beta = 1}^{k}$
of $M_k(\fA)$; we thus calculate
\begin{align}
  \cJ_{2;k}(\Eins_k \x T)
    &= \sum\nolimits_{i,j = 1}^{n} \sum\nolimits_{\alpha,\beta = 1}^{k}
       e_{\alpha\beta} \x E_{ij} \x [\Eins_k \x T](e_{\alpha\beta} \x E_{ij})\\
    &= \sum\nolimits_{i,j = 1}^{n} \sum\nolimits_{\alpha,\beta = 1}^{k}
       e_{\alpha\beta} \x E_{ij} \x e_{\alpha\beta} \x T(E_{ij}).
\end{align}
If $(f_p)_{p = 1}^{k}$ denotes the canonical basis of $\C^k$, any vector $x \in \cH_{A;k}$ may be written as
\mbox{$x = \sum_{p = 1}^{k} f_p \x x_p$} with elements $x_p \in \cH_A$; similarly, we write $y = \sum_{q = 1}^{k}
f_q \x y_q \in \cH_{B;k}$ with elements $y_q \in \cH_B$. By Lemma \ref{JamVar} we have to find conditions, such
that
\begin{equation}
  \SkM{\cJ_{2;k}(\Eins_k \x T)x \x y}{x \x y} \geq 0
\end{equation}
holds for all $x \in \cH_{A;k}$ and all $y \in \cH_{B;k}$. We write $x \x y = \sum_{pq} f_p \x x_p \x f_q \x y_q
= \sum_{rs} f_r \x x_r \x f_s \x y_s$ and rewrite the left hand side of this condition as
\begin{align}
  &\phantom{=\,\,} \sum\nolimits_{i,j,\alpha,\beta,p,q,r,s} \SkM{e_{\alpha\beta} f_p \x E_{ij} x_p \x e_{\alpha\beta}
    f_q \x T(E_{ij})y_q}{f_r \x x_r \x f_s \x y_s}_{\C^k \x \cH_A \x \C^k \x \cH_B}\\
  &= \sum\nolimits_{i,j,\alpha,\beta,p,q,r,s} \SkM{e_{\alpha\beta} f_p}{f_r}_{\C^k} \cdot \SkM{E_{ij} x_p}{x_r}_{\cH_A}
     \cdot \SkM{e_{\alpha\beta}f_q}{f_s}_{\C^k} \cdot \SkM{T(E_{ij})y_q}{y_s}_{\cH_B}\\
  &= \sum\nolimits_{p,q,r,s} \left[\sum\nolimits_{\alpha,\beta} \SkM{e_{\alpha\beta}f_p}{f_r}_{\C^k}
     \SkM{e_{\alpha\beta}f_q}{f_s}_{\C^k}\right] \SkM{\cJ_2(T)x_p \x y_q}{x_r \x y_s}_{\cH_A \x \cH_B}.\label{Jam12}
\end{align}
For the bracket due to $\SkM{e_{\alpha\beta}f_p}{f_r}_{\C^k} = \delta_{\alpha r} \delta_{\beta p}$ etc. we calculate
\begin{equation}
  \sum\nolimits_{\alpha,\beta} \SkM{e_{\alpha\beta}f_p}{f_r}_{\C^k} \SkM{e_{\alpha\beta}f_q}{f_s}_{\C^k}
  = \left(\sum\nolimits_\beta \delta_{\beta p} \delta_{\beta q}\right)
    \left(\sum\nolimits_\alpha \delta_{\alpha r}\delta_{\alpha s}\right)
  = \delta_{pq} \delta_{rs},
\end{equation}
which yields
\begin{equation}
  \sum\nolimits_{p,r = 1}^{k} \SkM{\cJ_2(T)x_p \x y_p}{x_r \x y_r}_{\cH_A \x \cH_B}
  = \SkM{\cJ_2(T)v}{v}_{\cH_A \x \cH_B},
\end{equation}
where $v = \sum_{p = 1}^{k} x_p \x y_p$. By Lemma \ref{JamSchmidt}, this vector $v$ has a Schmidt number not greater
than $k$, and this completes the proof. \BE\\
Note that in this proof we had to use $\cJ_2$; if we used $\cJ_1$ instead, the first term in the bracket
of~(\ref{Jam12}) would read $\SkM{e_{\alpha\beta}^*f_p}{f_r}_{\C^k} = \delta_{\alpha p}\delta_{\beta q}$,
which would lead to a different result. As an example, that the result does not hold for $\cJ_1$, consider
the case $\cH_A = \cH_B = \C^2$ and $\fA = \fB = \C^{2 \times 2}$, where $\Fkt{T}{\C^{2 \times 2}}{\C^{2 \times 2}}$
is the identity map, which obviously is completely positive. However, $v := e_1 \x e_2 - e_2 \x e_1 \in \C^2$
is an eigenvector of $\cJ_1(T)$ having eigenvalue $-1$, so that $\SkM{\cJ_1(T)v}{v}$ is negative. It has
Schmidt number $2$, so the analogue of Theorem~\ref{Main} with $\cJ_1$ instead of $\cJ_2$ is wrong.
\par We will conclude this section with two corollaries of the main theorem:
the first corollary is a special case of a more general result by Choi \cite{Ch}, the latter is known as the
Jamio{\l}kowski criterion for complete positivity (cf. e.\,g. \cite{Sal,KW}).
\begin{Corollary}[Complete positivity I]\hfill\\
  If a linear map $T \in L(\fA,\fB)$ is $\min\Mg{n,m}$-positive, it is also completely positive.
\end{Corollary}
\Proof By Theorem \ref{Main}, $T$ is $k$-positive, if and only if it is positive on all vectors having
Schmidt number not greater than $k$. Since any vector on $\cH_A \x \cH_B$ has Schmidt number not greater
than $\min\Mg{n,m}$, any $\min\Mg{n,m}$-positive map is completely positive. \BE
\begin{Corollary}[Complete positivity II]\hfill\\
  A linear mapping $T \in L(\fA,\fB)$ is completely positive, if and only if $\cJ_2(T)$ is positive semidefinite.
\end{Corollary}
\Proof If $T$ is completely positive, by Theorem~\ref{Main}, it must be positive on vectors having arbitrary
Schmidt number. Since any vector has a well-defined Schmidt number, $T$ must be positive semidefinite.
The converse statement is obvious. \BE

\section{Summary}\label{Summary}
\noindent We investigated the properties of two maps, $\cJ_1$ and $\cJ_2$, both of them being called
\anfEngl{Jamio{\l}kowski isomorphism}. We used Jamio{\l}kowski's criterion for positivity to give a concise
and simple proof of a relation between vectors having Schmidt number $k$ and $k$-positive maps, namely
that a map~$T$ is \mbox{$k$-positive}, if and only if $\cJ_2(T)$ is positive on vectors having Schmidt number not
greater than~$k$. Using this theorem we rederived the Jamio{\l}kowski criterion for complete positivity
and that $\min\Mg{n,m}$-positivity implies complete positivity.

\section{Acknowledgements}
\noindent The authors thank Gernot Alber for helpful discussions. K. S. Ranade is supported by a
graduate-student scholarship (Promotionsstipendium) of the Technische Universit\"at Darmstadt.
M. Ali acknowledges financial support by the Higher Education Commission, Pakistan, and the Deutscher
Akademischer Austauschdienst, Bonn.

\end{document}